\documentclass[letter]{aa} 

\usepackage{graphicx}
\usepackage{txfonts}
\begin{document} 

\title{The principle of maximum entropy explains the cores\\
    observed in the mass distribution of dwarf galaxies}
   \author{Jorge S\'anchez Almeida\inst{1,2}
     \and
     Ignacio Trujillo\inst{1,2}
     \and
     Angel Ricardo Plastino\inst{3}
      }
      \institute{Instituto de Astrof\'\i sica de Canarias, La Laguna, Tenerife, E-38200, Spain\\
        \email{jos@iac.es}
      \and
      Departamento de Astrof\'\i sica, Universidad de La Laguna
      \and
      CeBio y Departamento de Ciencias B\'asicas,  Universidad Nacional del Noroeste de la Prov. de Buenos Aires,  UNNOBA, CONICET, \\Roque Saenz Pe\~na 456, Junin, Argentina
             }
   \date{Received August 15, 2020; accepted \today}

 
   \abstract{
 Cold Dark Matter (CDM) simulations predict a central cusp in the mass distribution of galaxies. This prediction is in stark contrast with observations of dwarf galaxies which show a central plateau or {\em core} in their density distribution. The proposed solutions to this core-cusp problem can be classified into two types. Either they invoke feedback mechanisms produced by the baryonic component of the galaxies, or they assume the properties of the dark matter (DM) particle to depart from the CDM hypothesis. Here we propose an alternative yet complementary explanation.  We argue that cores are unavoidable in the self-gravitating systems of maximum entropy resulting from non-extensive statistical  mechanics. Their structure follows from the Tsallis entropy, suitable  for systems with long-range interactions. Strikingly, the mass density profiles  predicted by such thermodynamic equilibrium match the observed cores without any adjustment or tuning. Thus, the principle of maximum Tsallis entropy explains the presence of cores in dwarf galaxies.
}
   \titlerunning{Maximum entropy explains galaxy cores}
   \authorrunning{S\'anchez Almeida et al.}
   \maketitle
   %

\section{Introduction}\label{sec:intro}

The total mass density of low-mass galaxies flattens up at their center showing what is called a {\em core}. This observational fact was mentioned as a long standing problem of the $\Lambda$ CDM paradigm
(e.g., see the recent reviews by \citeauthor{2015PNAS..11212249W}~\citeyear{2015PNAS..11212249W} and \citeauthor{2017Galax...5...17D}~\citeyear{2017Galax...5...17D}),
since early DM-only numerical simulations predicted the existence of density cusps rather than cores in the inner regions of galaxies \cite[][]{1994Natur.370..629M}.  A popular explanation of the so-called {\em core-cusp problem} relies on including baryon physics in the simulations which, through gravity, couples baryon processes with DM.  Explosive baryon-driven events at the center of the galaxies produce sudden changes of the gravitational potential which, integrated over time, turn the DM distribution from cusp to core \cite[][]{2010Natur.463..203G}.  
Alternatively, the core-cusp problem may also point out a failure of the cold DM hypothesis \citep[][]{2015PNAS..11212249W,2017Galax...5...17D}.   Solutions include considering warm DM, so that its free-streaming velocities erase primordial fluctuations on small scales \citep{2000ApJ...542..622C}, or assuming self-interacting DM, so that the scattering between DM particles redistributes energy and momentum generating inner cores \citep[][]{2000PhRvL..84.3760S}.

Here we propose an alternative solution to the core-cusp problem based on the principle of maximum Tsallis entropy and the polytropes it leads to. For theoretical reasons presented in Sect.~\ref{sec:polytropes}, polytropes may provide a good representation for the  distribution of mass within galaxies, and they all have cores. Therefore, the question arises as to whether the cores of the polytropes reproduce the cores observed in the matter distribution of dwarf galaxies. Here we show that they do without any free parameter (Sect.~\ref{sec:result}). Polytropes describe thermodynamic (or meta-stable) equilibrium configurations of self-gravitating system under special conditions. Thus, our result suggests that these conditions are met in dwarf galaxies and may drive their internal structure (Sect.~\ref{sec:conclusions}).

\section{Maximum-entropy self-gravitating systems and polytropes} \label{sec:polytropes}
 Galaxies are  self-gravitating structures which, among all possible equilibrium configurations, choose only  those consistent with a stellar mass surface density  profile resembling a S\'ersic function \citep[e.g.,][]{2003ApJ...594..186B,2012ApJS..203...24V}\footnote{The S\'ersic functions include exponential disks, observed in dwarf galaxies \cite[e.g.,][]{1994A&AS..106..451D}, and {\em de Vaucouleurs} 1/4-profiles, characteristic of massive ellipticals \cite[e.g.,][]{1948AnAp...11..247D}.}. The settling into this particular configuration could be due to either some fundamental physical process (as it happens with the velocities of the molecules in a gas) or to the initial conditions that gave rise to the system \citep{2008gady.book.....B}. The mass distribution in galaxies is currently explained as the outcome of initial conditions \citep{2014ApJ...790L..24C,2015ApJ...805L..16N,2017MNRAS.465L..84L,2020MNRAS.495.4994B}. The option of a  fundamental process determining the configuration is traditionally discredited because, following the principles of statistical physics, it should correspond to the most probable configuration of a self-gravitating system and, thus, it should result from maximizing the entropy. Using the classical Boltzmann-Gibbs entropy leads to a distribution with infinity mass and energy  \citep{2008gady.book.....B,2008arXiv0812.2610P}, disfavoring this explanation. In the standard Boltzmann-Gibbs approach, however, the long-range forces that govern self-gravitating systems are not properly taken into account.
Systems 
with long-range interactions admit long-lasting meta-stable states described by a maximum entropy formalism based on Tsallis ($S_q$) non-additive entropies \citep[][and references there in]{1988JSP....52..479T,2009insm.book.....T}. Observational evidence for the $S_q$ statistics has been found in connection with various astrophysical problems \citep{2013SSRv..175..183L,2013ApJ...777...20S}. In particular, the maximization under suitable constraints of the Tsallis entropy of a Newtonian self-gravitating N-body system leads to a polytropic distributions \citep{1993PhLA..174..384P,2005PhyA..350..303L}, which
has finite mass and a shape closely resembling the DM distribution found in numerical simulations of galaxy formation \citep{2004MNRAS.349.1039N,2009PhyA..388.2321C}. 
In the current cosmological model, DM provides most of the gravitational pull needed for the ordinary matter to colapse forming visible galaxies, and thus, polytropes approximately describe the gravitational potential of galaxies.
As it is shown below, the mass density associated with a polytrope always has a core. The question arises as to whether the cores of the polytropes reproduce the cores observed in the matter distribution of dwarf galaxies, thus providing an alternative view for solving the core-cusp problem (Sect.~\ref{sec:intro}).

\begin{figure}
\centering 
\includegraphics[width=0.95\linewidth]{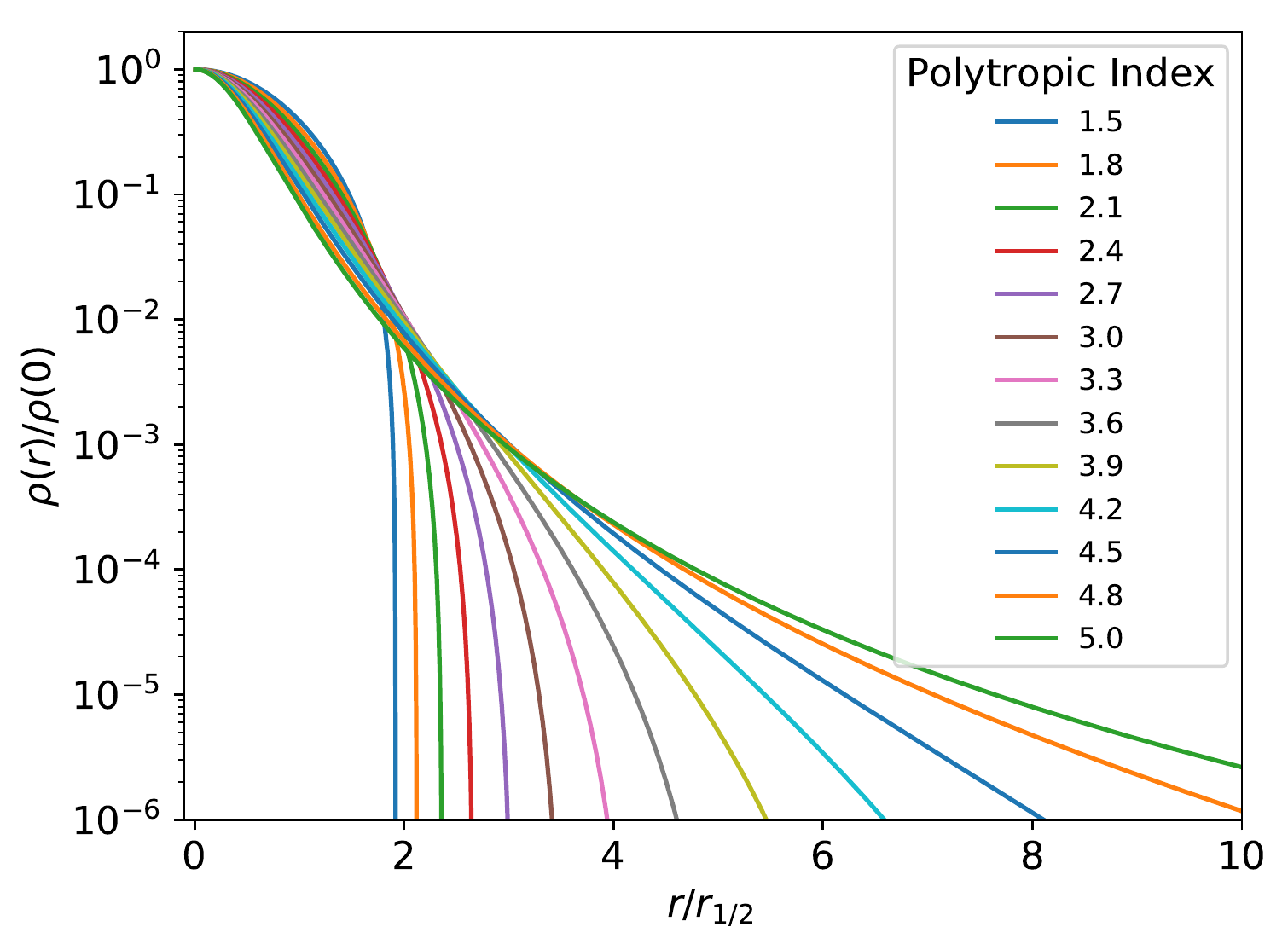}
\caption{
  Volume mass density resulting from the numerical solutions of the Lane-Emden equation (polytropes). Polytropes are  self-gravitating systems having maximum Tsallis entropy. The curves are normalized to the central density and to the half-mass radius ($r_{1/2}$). The examples show the range of physically plausible solutions, with the  corresponding polytropic index given in the inset.
}
\label{fig:lane_emden}
\end{figure}

A polytrope  of index $m$ is defined as the spherically-symmetric self-gravitating structure resulting from the solution of the Lane-Emden equation for the (normalized) gravitational potential $\psi$  \citep{1967aits.book.....C,2008gady.book.....B},
\begin{equation}
  \frac{1}{s^2}\frac{d}{ds}\Big(s^2\frac{d\psi}{ds}\Big)=
  \begin{cases}
    -3\psi^m & \psi > 0,\\
    0 & \psi \le  0.\\
  \end{cases}
\label{eq:lane_emden}
\end{equation}
 The symbol $s$ stands for the scaled radial distance in the 3D space and the mass volume density is recovered from $\psi$ as
 \begin{equation}
   \rho(r) = \rho(0)\,\psi(s)^m,
   \label{eq:densityle}
 \end{equation}
\begin{equation}
  r = b\, s,
   \label{eq:radius}
 \end{equation}
 where $r$ stands for the physical radial distance and $\rho(0)$ and $b$ are two arbitrary constants. Equation~(\ref{eq:lane_emden}) is solved under the initial conditions $\psi(0)=1$ and
 $d\psi(0)/ds =0$\,\,\footnote{Equation~(\ref{eq:lane_emden}) also admits solutions with $d\psi(0)/ds \not= 0$, but those are discarded because they have infinite central density and total mass \citep[e.g.,][]{2008gady.book.....B}.}.
 Figure~\ref{fig:lane_emden} illustrates the variety of physically admissible polytropes,  with the range of polytropic indexes
\begin{equation}
  3/2 \le m \le 5,
  \label{eq:nlimits}
 \end{equation}
 set because polytropes with $m\le 3/2$ are unstable or have infinite density  and those with $m > 5$ have infinite mass \citep{1993PhLA..174..384P,2008gady.book.....B}.
In order to compute the polytropes in Fig.~\ref{fig:lane_emden}, Eq.~(\ref{eq:lane_emden}) was split into a system of two first order differential equations for $\psi$ and $d\psi/ds$, which were integrated from $s=0$ using {\tt Lsoda} \citep{2019ascl.soft05021H} as implemented in {\em python} ({\em scipy.odeint}).

Note that all polytropes have cores, in the sense that $d\ln\rho/d\ln r\rightarrow 0$ when $r\rightarrow 0$. This property follows from the initial condition $d\psi(0)/ds=0$ and Eq.(\ref{eq:densityle}). It is shown by the density profiles displayed in Fig.~\ref{fig:lane_emden}.

\section{Results}\label{sec:result}

\begin{figure}
 \includegraphics[width=\linewidth]{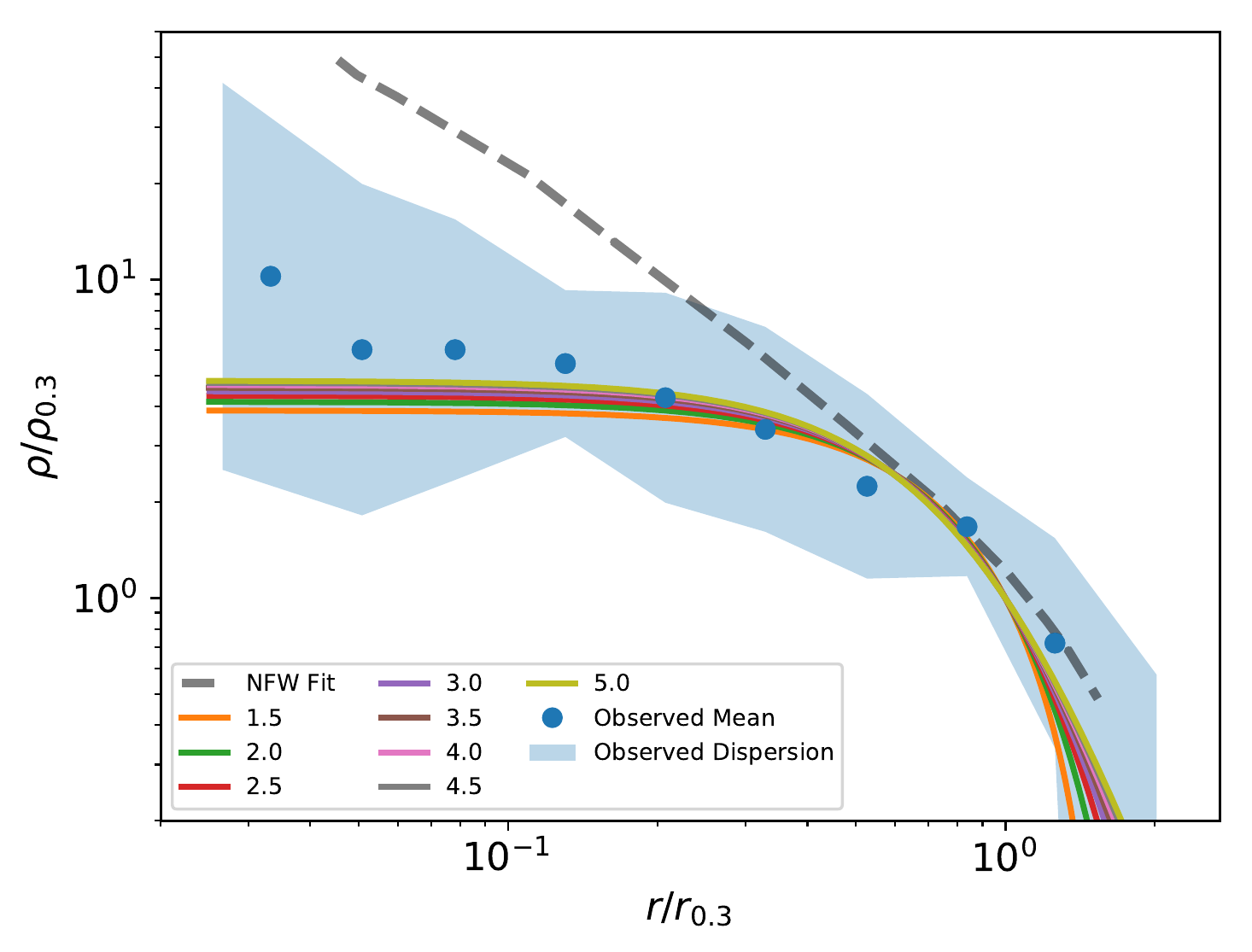}
 \caption{Density profile observed in the inner regions of the 26 {\em Little Things} galaxies by \citet{2015AJ....149..180O} (the blue symbols and the blue region give the mean and the RMS dispersion among the different objects). To reduce scatter, the observed densities and radii are normalized to the density and radius where the logarithmic derivative of the circular velocity equals 0.3 ($d\log v_c/d\log r =0.3$), denoted as $\rho_{0.3}$ and  $r_{0.3}$, respectively. Polytropes are parameter free in this representation (the solid lines with the corresponding indexes given in the inset). The dashed line gives a best-fit to the observed density using a NFW profile \citep{2015AJ....149..180O}, which does not follow the observed core.
}
  \label{fig:core-casp1}
\end{figure}
 Figure~\ref{fig:core-casp1} shows the state-of-the-art observation of galaxy cores in dwarf galaxies by \citet{2015AJ....149..180O}, which is based on 26 galaxies with  stellar masses $6.5 \le \log(M_\star/M_\odot) \le 8.2$  (blue symbols with the blue region giving the RMS dispersion among the different objects).
The total density is inferred from the circular-speed $v_c$ measured in the 21-cm hydrogen line which, for axi-symmetric systems, is  related to $v_c$ as \citep[e.g.,][]{2001ApJ...552L..23D},
  \begin{equation}
    \rho(r)= \frac{1}{ 4\pi G}\,\Big[\frac{v_c}{r}\Big]^2\,\Big[1+2\frac{d\log v_c}{d\log r}\Big],
    \label{eq:vc}
\end{equation}
were $G$ is the gravitational constant. The scatter of the 26 density profiles gets largely reduced when each individual profile is normalized  to the radius and density  where $d\log v_c/d\log r =0.3$, denoted as $r_{0.3}$ and $\rho_{0.3}$, respectively \citep{2015AJ....149..180O}. In addition to reducing the observational scatter, this normalization makes  the comparison with polytropes parameter-free. The density $\rho(r)$ consistent with a polytrope of index $m$  (Eq.~[\ref{eq:densityle}]) depends on two parameters $\rho(0)$ and $b$.   Using Eqs.(\ref{eq:radius}) and Eq.~(\ref{eq:vc}),
one can show that
  \begin{equation}
    \rho(x\,r_{0.3})\big/\rho_{0.3}=\psi^m(x\,s_{0.3})\big/\psi^m(s_{0.3}),
    \label{eq:scaling}
  \end{equation}
 where $x=r/r_{0.3}$  and $s_{0.3}$ is the value for $r_{0.3}$ obtained from $\psi(s)$. The right-hand side of Eq.~(\ref{eq:scaling}) does not depend on $\rho(0)$ or $b$, indicating that the same happens with the normalized density (the left-hand side of the equation) which, consequently, has no freedom in Fig.~\ref{fig:core-casp1}.
Thus, the agreement between the observed and the predicted cores is particularly revealing, suggesting a true connection between polytropes and the inner structure of dwarf galaxies.

  \section{Conclusions}\label{sec:conclusions}

We have shown that the polytropes, resulting from the principle of Tsallis entropy, reproduce  without any tuning the cores observed in the matter distribution of dwarf galaxies. The genesis of these cores is currently interpreted as driven by the interplay between baryons and DM, so that repetitive baryon motions modify the overall gravitational potential and the associated matter distribution  (Sect.~\ref{sec:intro}).  We note that the two explanations are not in contradiction. They are consistent if the baryon driven motions just shorten the time-scale needed to thermalize the global gravitational potential into a polytrope.  

Our study is focused on the central regions of the galaxies, but polytropes also work well in the outskirts.  The outer parts are fully dominated by DM, and it has been repeatedly shown that polytropes can be fit with Einasto profiles \citep[e.g.,][]{2006JCAP...06..008Z,2012MNRAS.423.2190S}, which fit well the outer parts of the DM profiles found in cosmological numerical simulations \citep[e.g.,][]{2004MNRAS.349.1039N,2005ApJ...624L..85M,2009PhyA..388.2321C}. In support of this, \citet{2015MNRAS.449.3645F} employ the maximum Tsallis entropy formalism to fit the radial dependence of $v_c$ in 24 galaxies with $8 \le \log(M_\star/M_\odot)\le 11$. $v_c(r)$ and $\rho(r)$ are interchangeable (Eq.~[\ref{eq:vc}]), so that the goodness of the fit at all radial distances also applies to $\rho(r)$ (even though \citeauthor{2015MNRAS.449.3645F} pay no specific attention to the cores studied here).

The association between dwarf galaxies and maximum Tsallis entropy opens up the possibility of using the well-proven tool-kit of statistical mechanics to understand them
\citep[][]{
  2008arXiv0812.2610P,
  2013MNRAS.430..121P,
  2013MNRAS.430.1578S}.
Identifying galaxies with polytropes  has a number of additional implications. Accurate mass profiles are needed to plan and interpret the astrophysical experiments to disclose the nature of DM.  DM annihilation cross-sections depend on halo shape \cite[e.g.,][]{2018PhRvD..97f3013Z}, and precise DM profiles and their time evolution should help us to distinguish between cold,  warm, or self-interacting DM \citep[e.g.,][]{2015PNAS..11212249W,2016MNRAS.460.1214L}.
The suite of mass models currently used in gravitational lensing studies does not include polytropes \cite[e.g.,][]{2001astro.ph..2341K}, but subtle details in the mass model are critically important when precise magnifications are needed, or when  lensing is used to derive cosmological parameters \citep{2001AJ....122..103K,2007JCAP...07..006E}.


The ability of polytropes to reproduce observed galaxy properties also has impact on the statistical mechanics side.
Comparison with the cosmic evolution of astronomical objects will shed new light on whether the $S_q$ entropies, besides providing H-funtionals able to select particular steady state solutions of the Vlasov equation \citep{2005PhyA..356..419C}, also have a deeper thermodynamical meaning for self-gravitating systems. 


\begin{acknowledgements}
 
%
%
Thanks are due to Jose Diego for discussions on the mass models used in gravitational lensing analysis,
to  Bruce Elmegreen for help and references on the stability of disk galaxies, and to Arianna Di Cintio for references on the {\em core-cusp problem}.
JSA acknowledges support  from the Spanish Ministry of Economy and Competitiveness (MINECO), project AYA2016-79724-C4-2-P (ESTALLIDOS).
IT also acknowledges financial support from the European Union's Horizon 2020 research and innovation programme under Marie Sk\l odowska-Curie grant agreement No 721463 to the SUNDIAL ITN network, from the State Research Agency (AEI) of the Spanish Ministry of Science, Innovation and Universities (MCIU) and the European Regional Development Fund (FEDER) under the grant with reference AYA2016-77237-C3-1-P, from IAC projects P/300624 and P/300724, financed by the Ministry of Science, Innovation and Universities, through the State Budget and by the Canary Islands Department of Economy, Knowledge and Employment, through the Regional Budget of the Autonomous Community, and from the Fundaci\'on BBVA under its 2017 programme of assistance to scientific research groups, for the project {\em Using machine-learning techniques to drag galaxies from the noise in deep imaging}.

\end{acknowledgements}

%




\end{document}